# An Atom-Pair Bond Theory for the Alloying of Metals

T. Rajasekharan[1,*] and V. Seshubai[2,**]

[1]*Defence Metallurgical Research Laboratory, Kanchanbagh P.O. Hyderabad 500 058, India,*

[2]*School of Physics, University of Hyderabad, Hyderabad 500 019, India.*

We present an Atom-Pair Bond (APB) theory for the energy of a metallic bond based on the ideas of covalent bonding proposed by Pauling. An expression is derived which accurately predicts the signs of the heats of formation of binary alloys. It also explains the characteristics of the Rajasekharan-Girgis lines and their ability to predict accurately concomitant and mutually exclusive structure types in phase diagrams. Evidence is provided to show that the charge transfer on the atom-pair bond which is central to the present theory decides the experimentally observed volume changes on alloying.

Pauling had proposed that a combination of s, p and d orbitals can form suitable bonding orbitals in metals [1]. He had defined a set of valences for metals [1, 2] which could qualitatively explain many properties of transition metals. Pairs of electrons, occupying bond positions between adjacent pairs of atoms, carry out unsynchronised or partially synchronised resonance throughout the crystal. The shortest bonds would contribute maximum to the resonance and the longer ones to a lesser extent. Pauling suggested that unsynchronised resonance would not only increase the stability of the metallic lattice, but would also give a simple explanation to electrical and thermal conductivity of metals [3]. However, the difficulties in making RVB calculations [4] and also the widespread belief from long time ago [5] that metals cannot be covalently bonded have prevented extensive use of the theory.

Metals are characterised by high ligancy ($\geq 8$) and high symmetry [6]. The first nearest A-B bonds which are several due to high ligancy can be expected to play a significant role in deciding the energy of an alloy. We hypothesise that the nearest-neighbour A−B bond energy (and distance) is nearly the same in compounds occurring at different compositions in the same binary system. This assumption is similar to the idea in conventional chemistry that the bond energies for a particular type of bond, say an S−H bond, is found to be approximately constant in different molecules [7] containing that type of bond. Pauling considered [8] charge transfer in alloy systems and used an assumption along similar lines to arrive at the volume of an A−B ion-pair in the binary system Ca−Pb with the compounds $Ca_2Pb$, $Ca_5Pb_3$, CaPb (tP4), CaPb (cP4) and $CaPb_3$. He assumed that after equal number of Ca and Pb ions form bonds, the extra Ca and Pb atoms have the effective volumes of the elementary substances (i.e. they are not affected by the bonding process). Using the observed values of the mean atomic volume in the alloy, he then calculated the volumes of the $Ca^+ - Pb^-$ ion pair in all the intermetallic compounds in the Ca−Pb system and found them to be nearly the same. Similar observations were reported by him [8] in the case of Co−Zr and Co−Ga systems.

We have verified the above for all the 45 binary systems in which $MgCu_2$ type and $CaCu_5$ type compounds coexist [9]. The ratio of the shortest A−B bond lengths is close to 1.05. Such an effect is also demonstrated for all the 33 binary systems in which $MgCu_2$ and CsCl type compounds co-exist [9].

Consider two atoms A and B of electronegativity $\chi_A$ and $\chi_B$ and valence $N_A$ and $N_B$, with $\chi_A > \chi_B$. The atom with the higher electronegativity has often the higher valence [10], so $N_A > N_B$. We now make an important assumption whose validity is sought to be borne out by its ability to predict accurately the outcome of experiments. Due to the high ligancy in metals [6] and due to resonance between various available bond positions, we approximate that the valence of a metallic atom effective in the formation of a linear bond is $N^{1/3}$.

The bond formed between A and B will have a certain ionic character and the increase in stability of the heteronuclear bond A−B in comparison with the average of the homonuclear bonds can be attributed to the ionicity in the bonds. The negative contribution to the bond energy due to this effect is equal to [11, 12]

$$\Delta H_{-ve} = -23\ (\chi_A - \chi_B)^2\ \text{kcal/mol} = -(\chi_A - \chi_B)^2\ \text{eV/bond.} \qquad \dots (1).$$

Pauling pointed out [8] that in metallic compounds, maximum stability would result from equalising the number of valence electrons on A and B, thus maximising the number of bonds between A and B, by a transfer of electrons from A to B. Thus the number of electrons/atom on the atom-pair bond after electron transfer would be $\left(N_A^{1/3} + N_B^{1/3}\right) / 2$. The number of electrons transferred would then be $\left(N_A^{1/3} - N_B^{1/3}\right) / 2 = \left(\Delta N^{1/3}/2\right)$.

The ionicity of the bond deposits extra negative charges near the more electronegative atom A, making the atom B positively charged. Pauling has pointed out that the electroneutrality principle would necessitate [8] charge transfer from A to B so that there are no net positive charges on B. We therefore infer that the charge transfer (i.e. $\Delta N^{1/3}/2$) from A to B has to increase with an increase in the electronegativity difference ($\Delta\chi$) between the atoms on the metallic bond.

The transfer of electrons is from an atom with higher electronegativity $\chi_A$ to one with lower electronegativity $\chi_B$ and would make a positive contribution to the bond energy. The energy involved in removing an electron from an atom [13, 14] is $\frac{I_{atom} + E_{atom}}{2}$ and hence the positive contribution is

$$\Delta H_{+ve} = \left\{\left(\frac{I_A + E_A}{2}\right) - \left(\frac{I_B + E_B}{2}\right)\right\} . \frac{\Delta N^{1/3}}{2} \qquad \ldots(2).$$

$I_A$ and $E_A$ in electron volts are the first ionisation energy and electron affinity respectively of atom A, and $I_B$ and $E_B$ are those of atom B. $\Delta H_{+ve}$ is the difference in energy in eV, involved in the removal of $(\Delta N^{1/3}/2)$ electrons from the more electronegative atom A and its subsequent absorption by atom B.

We use the equation $\chi = 0.187\,(I_{atom} + E_{atom}) + 0.17$ from [15], connecting Pauling's electronegativity values ($\chi$) and $\frac{I_{atom} + E_{atom}}{2}$ to write the total energy of the atom pair bond, with $\Delta\chi = (\chi_A - \chi_B)$, as

$$\Delta H_{Bond} = \Delta H_{-ve} + \Delta H_{+ve} = -(\Delta\chi)^2 + 1.34\,\Delta\chi\,\Delta N^{1/3} \quad \text{in eV} \qquad \ldots(3).$$

Following the early work of Axon [16] and Mott [17], Miedema et al. [18-20] had developed an empirical equation that predicts the signs of the heats of formation ($\Delta$H) of equiatomic compounds in binary metallic systems as given below

$$\Delta H = \left[-\Delta\phi^2 + \frac{Q}{P}\left(\Delta {N_{ws}}^{1/3}\right)^2 - \frac{R}{P}\right] \quad \ldots(4).$$

$\phi$ and $N_{ws}$ are the work function and the 'electron density at the boundary of the Wigner-Seitz cells' of the elements. The numerical values of $\phi$ and $N_{ws}$ were obtained from Pauling's electronegativity and the experimental bulk modulus of the elements respectively. Their values were adjusted within uncertainties in their determination to predict the signs of $\Delta$H of more than 500 binary systems, using Eq. (4), with nearly 100% accuracy. Considerable amount of empiricism was involved in developing Miedema's equation. It was reported that the choice of $\left(\Delta\, {N_{ws}}^{1/3}\right)^2$ in Eq. (4), rather than $(\Delta N_{ws})^2$ which followed from Miedema's theoretical arguments, gave a better fit to the experimental data [19], with P and Q then becoming universal constants for all elemental combinations, with Q/P = 9.4 Volts/ (density units) 1/3. The value of $Q/P$ was obtained from a fit of the experimental information on signs of $\Delta$H. The term $R/P$, increasing with the valence of the p-metals, was introduced to correctly reproduce the empirically observed data on the signs of $\Delta$H of transition metal (t) - p-metal combinations. $R/P$ is zero for other combinations. Miedema et al. suggested that $R/P$ has its origin in '$d-p$ hybridisation' [18]. Eq. (4) was multiplied by a function f(c) to arrive at the numerical values of $\Delta$H for compounds at different stoichiometries [19]. The basic form of f(c) was chosen so as to reproduce the experimentally observed general trends in the variation of $\Delta$H with concentration [19]. Though the signs of $\Delta$H were predicted very accurately by Eq. (4), the predicted magnitudes of $\Delta$H were much less accurate [21].

The electronegativity values for metals $\chi_M$ can be obtained from the linear relationship [18] between the corrected $\phi$ values and Pauling's electronegativity values.

$$\chi_M = (\phi - 0.662)/\ 2.157 \quad \ldots(5).$$

Such $\chi_M$ values are not far from the original electronegativity values of Pauling for the elements, and can be considered appropriate for use with metals due to the corrections introduced.

We also see from Fig 1 (a) that in the first long period of the periodic table, Pauling's valences for metals $N$ are comparable in magnitude, and vary in the same way with the atomic number of elements, as do Miedema's $N_{ws}$. From Fig. 1 (b), we see that $N_{ws}$ follows the variation in the melting and boiling points of elements, including the anomalous behaviour of Mn, more closely than Pauling's valences. Similar dependencies can be demonstrated for elements of the other periods of the periodic table as well. We recall that$N_{ws}$ values, which are close to Pauling's valences for metals, too had been adjusted to predict the signs of $\Delta H$ accurately. We therefore choose the values of $N_{ws}$ as the accurate valences $N$ for metallic elements.

It follows from Eq. (3) that the energy of the atom pair bond will be negative if $\Delta\chi / \Delta N^{1/3} > 1.34$. Eq. (3) states, when $\chi$ is replaced by $\phi$ using Eq. (5), that $\Delta H_{Bond}$ will be negative when $\Delta\phi / \Delta N^{1/3} > 2.9$. This relationship derived from the present Atom-Pair Bond (APB) theory agrees well with the condition found empirically by Miedema et al. [17], that a line of slope $\sqrt{Q/P} = 3.07$ separates the compounds with negative and positive heats of formation on a $\left( |\Delta\phi|, |\Delta N^{1/3}| \right)$ map. Using $\chi_M$ and $N$ values for elements derived from Miedema's corrected parameters, Eq. (3) predicts the signs of the heats of formation of metallic elements with the same high accuracy as does Miedema's equation. Identification of Eq. (3) as the energy of an $A - B$ bond also tells us why Miedema's equation predicts the *signs* of the heats of formation of both *solid and liquid alloys* with the same high accuracy [20], long range order has no role in deciding the energy calculated using Eq. (3).

We also note that the $R/P$ term of Eq. (4) for $t - p$ compounds follows from the theory proposed above. Pauling had defined [8] hyperelectronic elements, buffer elements and hypoelectronic elements. The hyperelectronic elements (mostly $p$ −metals) have more electrons than there are orbitals and hence have electrons paired up in their orbitals. According to the present theory, an amount of charge equal to $\Delta N^{1/3}/2$ would be transferred from the higher electronegativity atom, i.e. the $p-$metal atoms in the case of $t-p$ combinations, to the lower electronegativity atoms. This would generate from lone-pair breaking, an additional number of valence electrons on the $p-$metal atoms equal to $\Delta N^{1/3}/2$, leading to additional stability. This mechanism accounts for the $(R/P)$ term proportional to the valence of p metals found necessary by Miedema et al., to correctly predict the signs of the heats of formation of $t-p$ combinations.

In Fig. 2, we show the intermetallic compounds crystallising with the $SiCr_3$ type crystal structure, with the $t-p$ compounds marked red (open circles), on a map using $\Delta\chi_M$ and $\Delta N^{1/3}$ as coordinates. The list of compounds crystallising with $SiCr_3$ type structure is from the literature [22-24]. The $t-p$ compounds which are stabilised by the lone$-$pair breaking mechanism according to the present theory occur in the region where the $A-B$ bond is predicted to have a positive heat of formation by Eq. (3).

In Fig. 3, we show a plot of $\Delta H_{EXPT}$ versus $\Delta H_{Bond}$ calculated using Eq. 3, for the $MgCu_2$ type compounds. The experimental values of the heats of formation of $MgCu_2$ type compounds $\Delta H_{EXPT}$ were obtained from a recent review [25]. The regression factor of the linear fit is 0.87. The heats of formation of new compounds of the above structure type can be obtained by interpolation from the figure. We note that there are no empirically derived constants in Eq. (3).

Eq. (3) calculates the energy of the nearest neighbour $A-B$ bond only. The calculation of the heat of formation of a compound from that of the bond needs accurate input on the relative contributions of different sets of bonds to the total energy. Resonance can also be expected to alter the energy of the compound.

Rajasekharan and Girgis [26] had reported that binary systems with compounds belonging to a particular structure type fall on a straight line on a $(\Delta\phi, \Delta N^{1/3})$ map. Rajasekharan-Girgis (RG) lines of some structure types occurring at composition $AB_3$, together with that of the $MgCu_2$ type occurring at composition $AB_2$, are shown in Fig. 4. Depending on which straight lines pass through the point corresponding to a given binary system on the map, one can predict the concomitant and mutually exclusive structure types in binary systems with great accuracy. The figure tells, for instance, that if an $MgCu_2$ type compound occurs at the composition $AB_2$, an $AsNa_3$ type (or $CoGa_3$ type or $SiCr_3$ type) compound will not occur at composition $AB_3$. It is surprising that predictions regarding the structures adopted by intermetallic compounds can be made using Miedema's parameters because 'structural energies' are considered small and Miedema's theory isotropic. No theoretical model in the literature for the alloying of metals anticipates such effects. One important result of identifying Eq. (3) with the energy of the A−B bond is that each point on the $(\Delta\chi,\ \Delta N^{1/3})$ map corresponds to a single value of energy for the A−B bond. We see that the electroneutrality principle would require $\Delta\chi$ (and therefore, $\Delta\phi$) to be proportional to $\Delta N^{1/3}$ on the bond, giving rise to straight lines on the Rajasekharan-Girgis (RG) map for compounds of a particular structure type. All the RG lines corresponding to structure types concomitant in a binary system would then pass through the point corresponding to the energy of the A−B bond in that system, thus explaining the ability of the RG maps to predict concomitant and mutually exclusive structure types in metallic binary systems.

In further support of the conclusion that $\Delta N^{1/3}/2$ represents charge transfer between atoms on alloy formation, in Figs. 5(a) and (b), we show the volume changes (in %) of $MoSi_2$ and $MgCu_2$ type compounds on alloy formation, as a function of $\Delta N^{1/3}$. We recall that transfer of electrons to an atom causes an increase in its valence and reduction of its size, except in the case of the buffer elements [8]. We observe that the volume changes go to zero at $\Delta N^{1/3} = 0$ in both cases and increase systematically with increase of $|\Delta N^{1/3}|$ supporting the conclusion that $\Delta N^{1/3}/2$ represents the charge transfer on alloy formation.

TR thanks DMRL, Hyderabad, for permission to publish this paper. VS thanks UGC for research funding under UPE and CAS programs.

*trajasekharan@gmail.com , ** seshubai@gmail.com

[1] L. Pauling, Phys. Rev. 54, 899 (1938).

[2] L. Pauling, The Nature of the Chemical Bond, 3rd edition (Oxford and IBH publishing, 1975) p. 93.

[3] L. Pauling , Jl. Solid state Chem. 54, 297 (1984).

[4] J. R. Mohallem, R.O. Vianna , A.D. Quintao, A. C. Pavao and R. McWeeny, Z. Phys. D **42**, 135 (1997).

[5] W. L. Bragg, Jl. Roy. Soc. Arts **85**, 431 (1937).

[6] F. Laves, in *Intermetallic compounds,* ed. Westbrook J. H. (John Wiley and Sons, Inc., New York, 1966), p.129-143

[7] W.L. Jolly, *Modern inorganic chemistry, 2nd ed*., (McGraw-Hill, Inc., N.Y. ,1991), p. 61

[8] L. Pauling, Proc. Natl. Acad. Sci. U.S.A. **84**, 4754 (1987).

[9] See EPAPS Document No. [Encl. 1] for details.

[10] L. Pauling, Proc. of Natl. Acad. Sci. **36**, 533(1950).

[11] L. Pauling, *The nature of the chemical bond, 3rd edition* (Oxford and IBH publishing ,1975), p. 92

[12] W. L. Jolly, *Modern inorganic chemistry, 2nd ed.* (McGraw-Hill, Inc., N.Y. 1991) p. 62

[13] R.S. Mulliken, J. Chem. Phys. **2**, 782 (1934).

[14] R.S. Mulliken, J. Chem. Phys., **3**, 573 (1935).

[15] J. E. Huheey, *Inorganic chemistry, 2nd edition* (Harper & Row, New York, 1978), p.167.

[16] H. J. Axon, Nature **162**, 997 (1948).

[17] B. W. Mott, Phil. Mag. **2**, 259 (1957).

[18] A. R. Miedema, P. F. de Chatel and F. R. de Boer, Physica **100B**, 1(1980).

[19] A. R. Miedema, R. Boom and F. R. de Boer, Jl. Less-Common Metals **41**, 283(1975).

[20] R. Boom, F. R. de Boer and A. R. Miedema, Jl. Less-Common Metals **45**, 237-245 (1976).

[21] C. Xing-Qiu, W. Wolf, R. Podloucky and P. Rogl, Intermetallics **12**, 59-62 (2004)

[22] W. B. Pearson, *A Handbook of Lattice Spacings and Structures of Metals and Alloys* (Pergamon Press, Oxford, 1967)

[23] P. Villars and L. D. Calvert, *Pearson's Handbook of Crystallographic Data for Intermetallic Phases*, (American Society for Metals, Metals Park, OH 44073, 1985)

[24] Binary alloy phase diagrams, 2nd edition plus updates on CD, Ed. In Chief T. B. Massalski (ASM International, Materials Park, OH 44073, 1990)

[25] J. H. Zhu, C. T. Liu, L. M. Pike and P. K. Liaw, Intermetallics 10, 579 (2002).

[26] T. Rajasekharan and K. Girgis, Phys. Rev. B 27, 910 (1983).

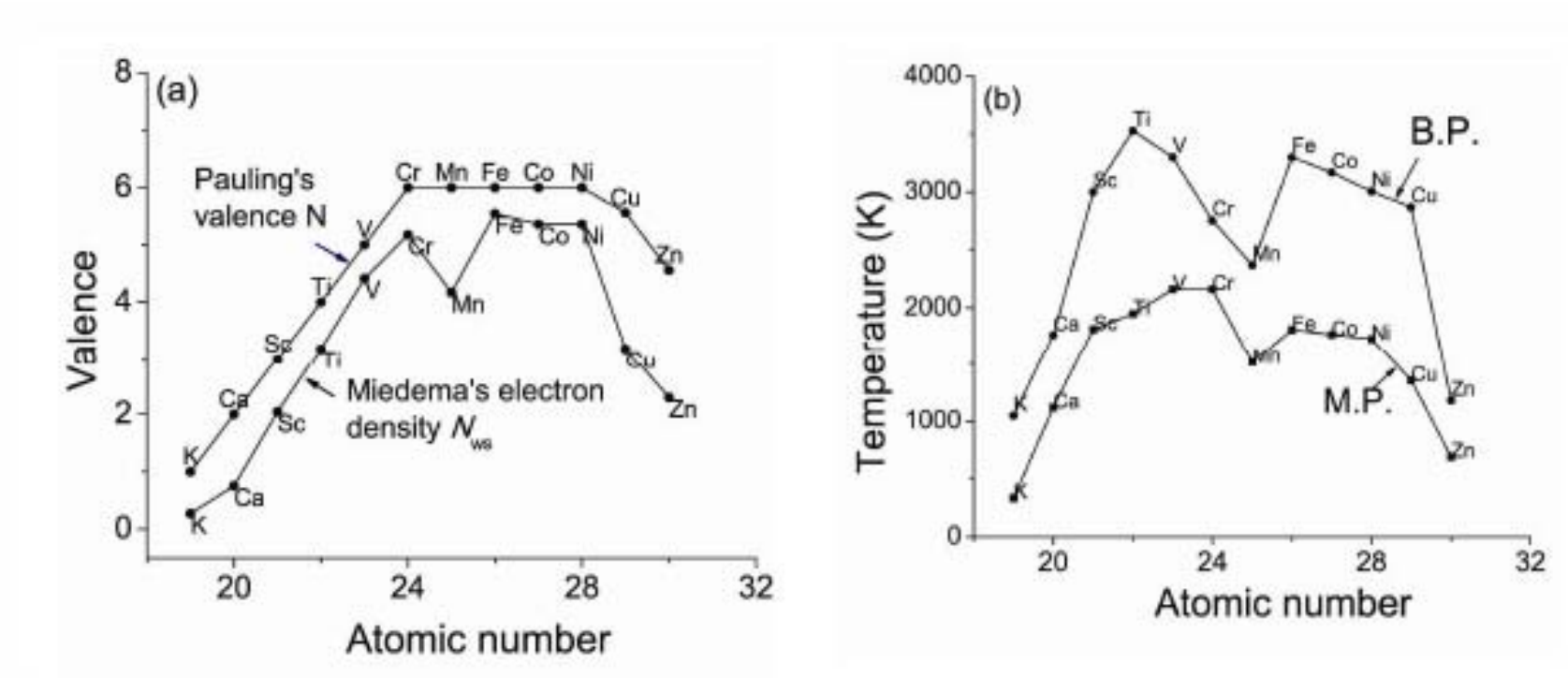

**FIG**. 1. **(a)** Pauling's valence for metals and Miedema's electron density parameter $N_{ws}$ are plotted versus the atomic number in the first long period. **(b)** The melting and the boiling points of elements in the first long period are plotted versus the atomic number of the elements. We see that Miedema's parameters whose values were obtained from the bulk modulus values and corrected to accurately predict the signs of the heats of formation of binary metallic systems are comparable to Pauling's valences and closely follow the variation of the melting and boiling points of the elements. The numerical values of $N_{ws}$ can be adjusted by a constant value to match Pauling's valences even better, but we have not attempted it herein since only the difference in $N_{ws}$ of the elements are relevant to our (or Miedema's) conclusions.

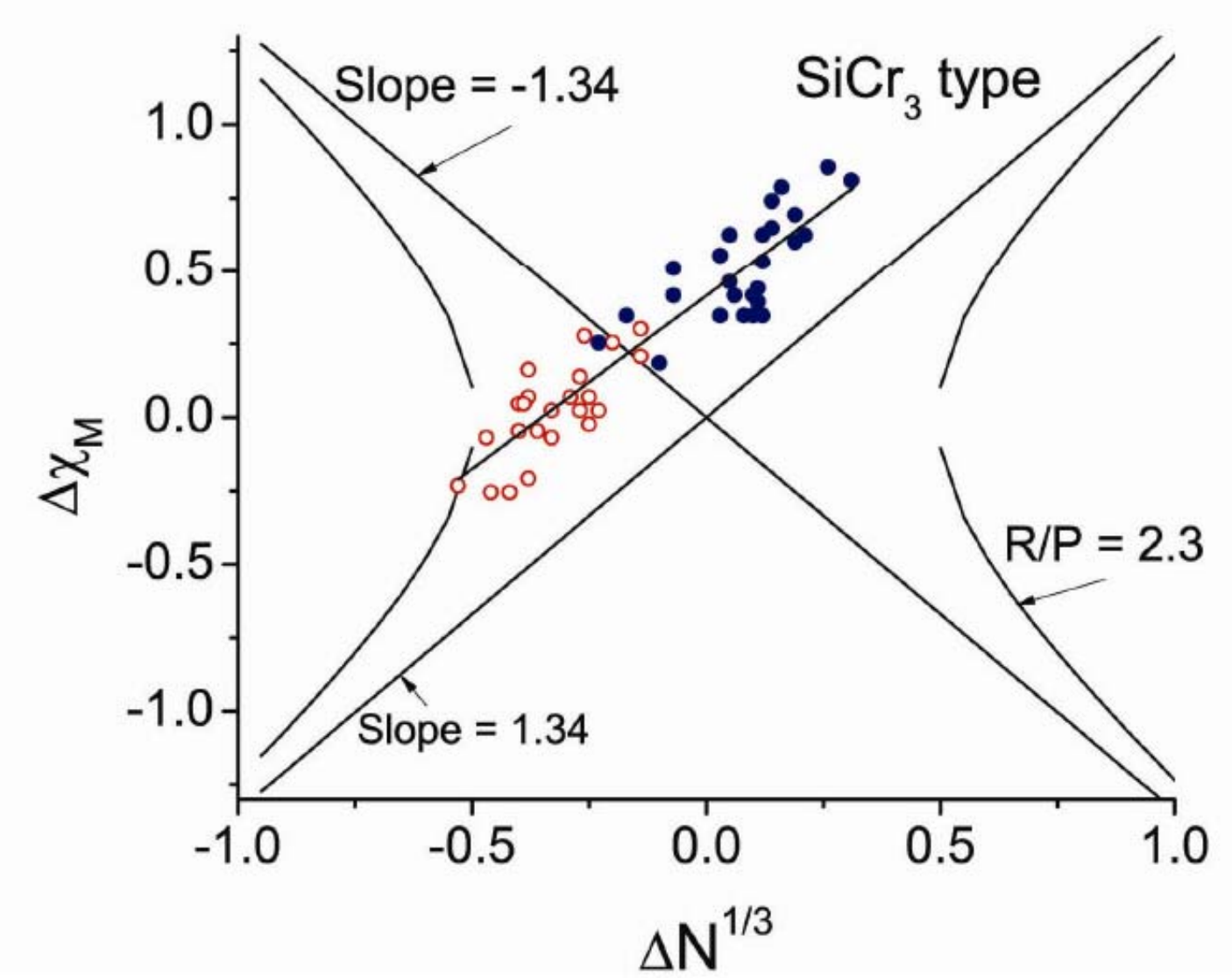

FIG. 2 (color online). The compounds belonging to the $SiCr_3$ structure type are shown on a $(\Delta\chi,\ \Delta N^{1/3})$ map. Each point on the map stands for a binary system. The compounds of hyperelectronic p− metals (open circles, red), have an additional stabilizing mechanism due to lone-pair breaking. They are seen to occur in a region otherwise predicted to have positive bond energy by Eq. (3). The hyperbola corresponds to Miedema's R/P = 2.3 which is equivalent to an additional stabilization of 0.49 eV.

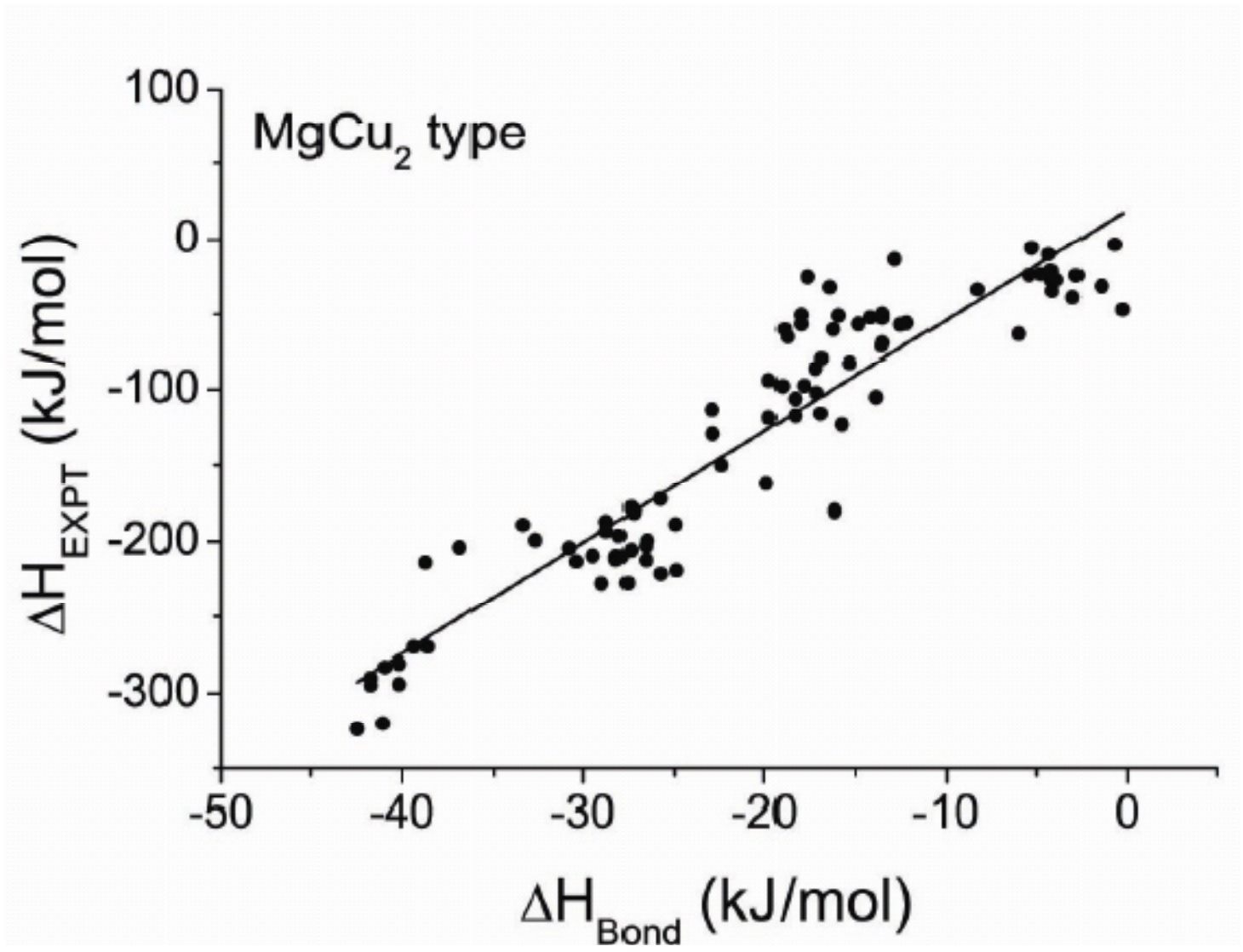

FIG. 3. The experimentally determined heats of formation $\Delta H_{EXPT}$ (kJ/mol) of $MgCu_2$ type compounds is compared with the bond energy $\Delta H_{Bond}$ calculated using Eq. (3). The energy obtained in eV/bond from Eq. (3) is multiplied by 96.485 to convert it into units of kJ/mol.

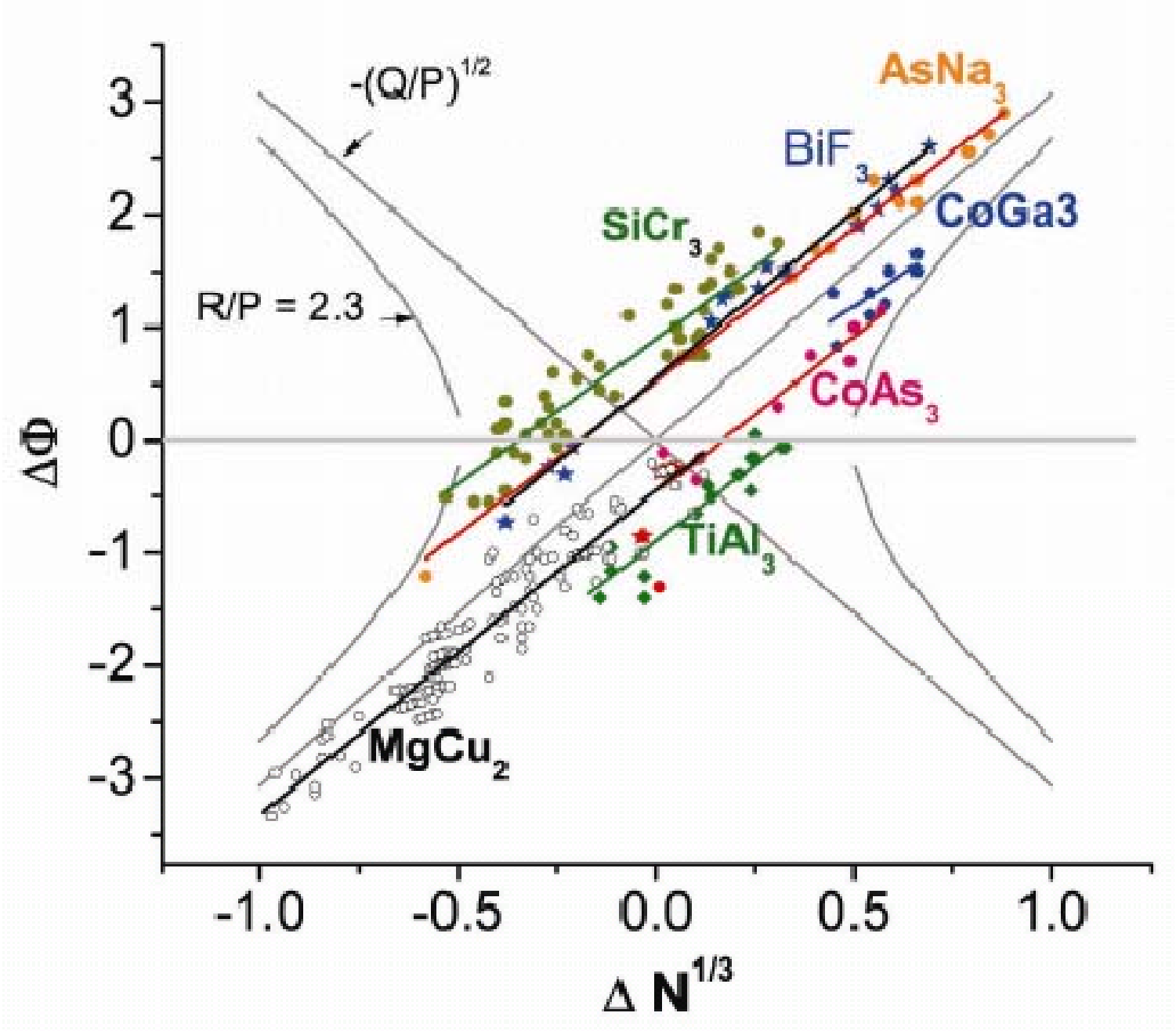

FIG. 4 (color online). Examples of Rajasekharan-Girgis (RG) lines of some structure types occurring at the stoichiometry $AB_3$ are shown along with that of the $MgCu_2$ type occurring at composition $AB_2$. We see that they all form straight lines with positive slopes nearly equal to that of the line that demarcates the regions of positive and negative heats of formation. By observing whether a binary system occurs on the line corresponding to a structure type or not, it is possible to predict whether that structure type can occur in the binary system. Also the structure types coexisting in a particular binary system have their RG lines passing through the point corresponding to that binary system. For instance, we can predict from the above figure that if an $MgCu_2$ type compound occurs in a binary system at composition $AB_2$, an $SiCr_3$ type compound will not occur at composition $AB_3$. In the overlap region of $AsNa_3$ type and $BiF_3$ type, the compounds $BiK_3$, $BiRb_3$ and $SbLi_3$ show structural transformations between the two structure types as a function of temperature [24].

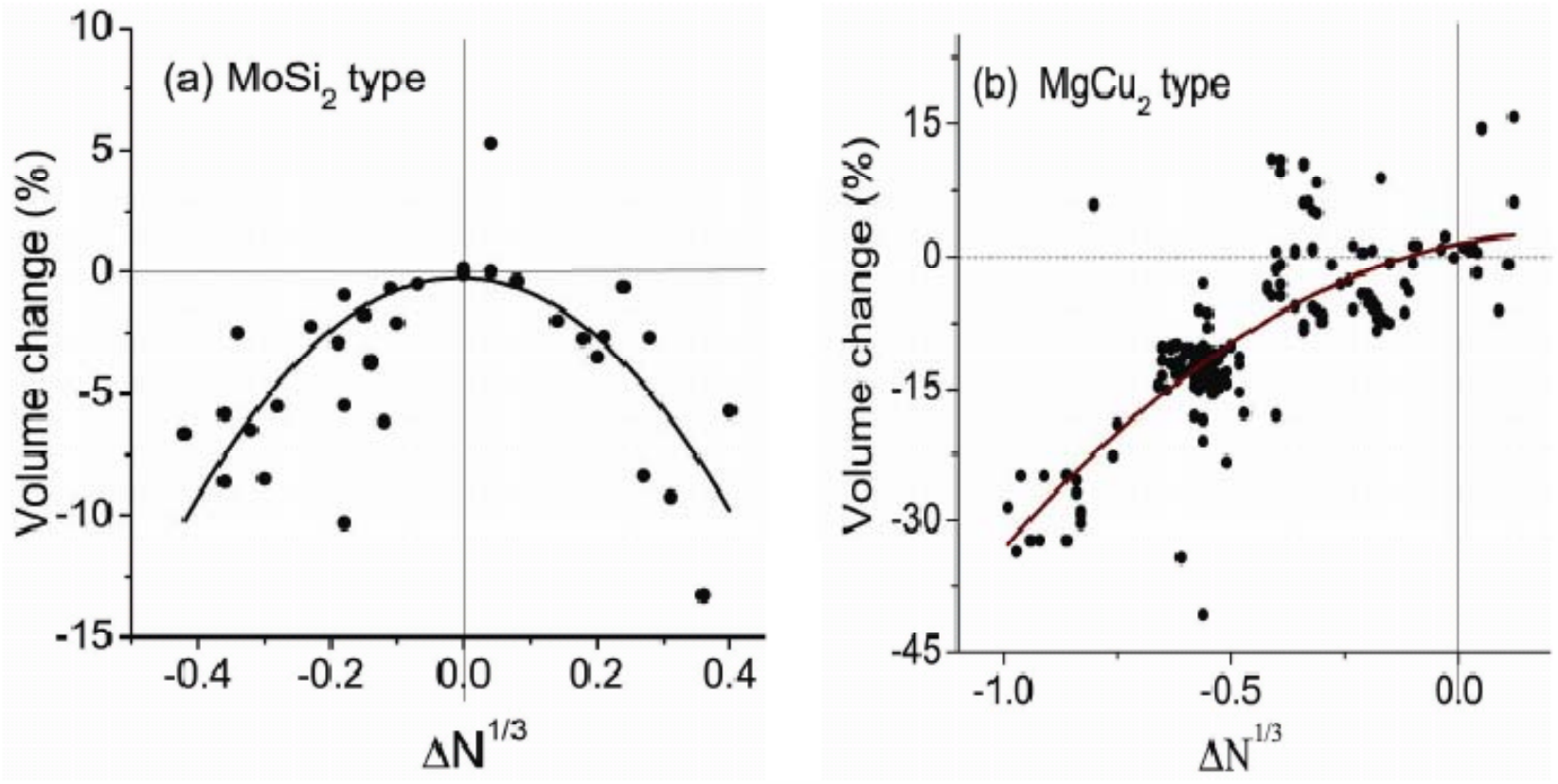

Fig. 5. Percentage change in volume on alloying as a function of $\Delta N^{1/3}$, (a) for $MoSi_2$ type compounds and (b) for $MgCu_2$ type compounds. The volume per molecule estimated from the molar volumes of the elements is subtracted from the volume per molecule computed from the unit cell volumes of the compounds, and expressed as a percentage relative to the volume per molecule before alloying.

EPAPS Document No. [Encl.1]

# An Atom-Pair Bond Theory for the Alloying of Metals

**T. Rajasekharan* and V. Seshubai**

**For correspondence:** trajasekharan@gmail.com, Seshubai@gmail.com

## Comparison of nearest neighbour A−B bond distances in concomitant compounds:

We initially choose $MgCu_2$ type ($AB_2$) and $CaCu_5$ type ($AB_5$) compounds as examples. The choice is governed by the fact that there are a large number of compounds crystallising in both the structure types and that the $CaCu_5$ type compounds are at a far off stoichiometry with 3 more B atoms than the $MgCu_2$ type compounds. Out of the 181 binary systems with $MgCu_2$ type compounds and the 79 binary systems with $CaCu_5$ type compounds, 43 binary systems have compounds with both the structure types. Out of the 224 binary systems with CsCl type (AB) compounds, 33 have $MgCu_2$ type compounds at $AB_2$. The data are from [1-3].

The crystal structures of the $MgCu_2$ and $CaCu_5$ type compounds are shown in Figures 1 (a) and (b). The shortest A−B distances are marked.

The CsCl structure type is bcc with the cube centre to the corner as the shortest A−B distance. In Tables I and II, we compare the nearest-neighbour bond lengths in $MgCu_2$ type compounds with those in concomitant $CaCu_5$ type and CsCl type compounds respectively. We observe that the ratio of the bond lengths is nearly constant and close to 1. Thus we see that the nearest neighbour A−B bond which can contribute most to the resonance and is the most relevant in deciding the energy of the compound, has nearly the same length in the compounds occurring at different compositions in the same binary system.

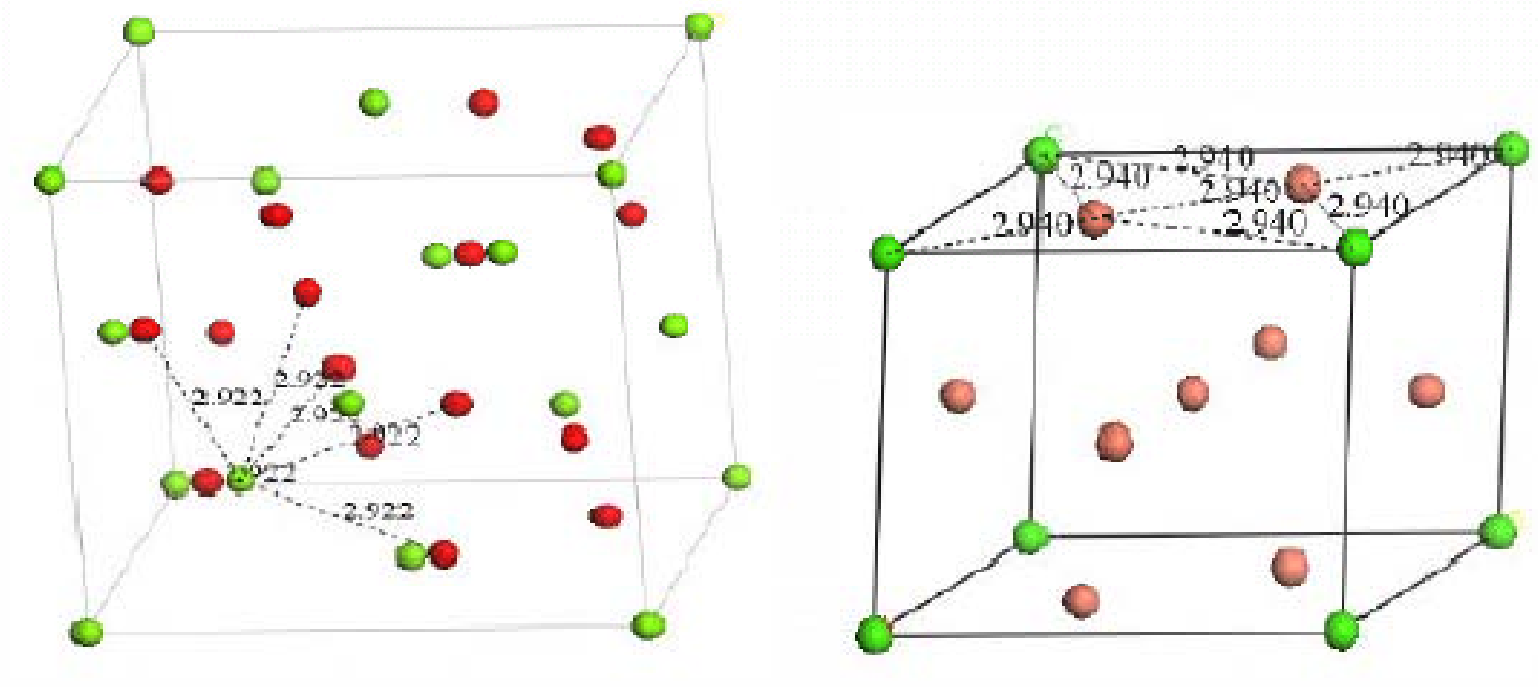

(a) (b)

FIG. 1. (a) The unit cell of $MgCu_2$ type ($AB_2$, cF24) compounds. A atoms are shown in green color. There are 12 bonds of equal length to the B atoms from each A atom. In the compound $MgCu_2$, the Mg-Cu nearest neighbor distance is 2.922 Å [4]. (b) The unit cell of $CaCu_5$ type ($AB_5$, hP6) compounds. A atoms are shown in green color. There are 6 bonds of equal length to the B atoms from each A atom. In the compound $CaCu_5$, the Ca-Cu nearest neighbor distance is 2.940 Å.

## References

[1] W. B. Pearson, *A Handbook of Lattice Spacings and Structures of Metals and Alloys* (Pergamon Press, Oxford, 1967).

[2] P. Villars and L. D. Calvert, *Pearson's Handbook of Crystallographic Data for Intermetallic Phases*, (American Society for Metals, Metals Park, OH 44073, 1985).

[3] *Binary alloy phase diagrams, 2nd edition plus updates on CD*, Ed. In Chief T. B. Massalski (ASM International, Materials Park, OH 44073, 1990).

[4] W. B. Pearson, *The Crystal Chemistry and Physics of Metals and Alloys* (Wiley-Interscience, New York, 1972) p. 655.

**TABLE I. The shortest bond lengths (d in Å) in the $MgCu_2$ type and $CaCu_5$ type compounds**

| Binary system | $CaCu_5$ type a (Å) | c(Å) | $d_1$(Å) | $MgCu_2$ tpe a (Å) | $d_2$ (Å) | RATIO $d_2/d_1$ |
|---|---|---|---|---|---|---|
| Ba-Pd | 5.49 | 4.34 | 3.170 | 7.95 | 3.296 | 1.04 |
| Ba-Pt | 5.51 | 4.34 | 3.181 | 7.92 | 3.284 | 1.03 |
| Ca-Ni | 4.93 | 3.93 | 2.846 | 7.26 | 3.010 | 1.06 |
| Ce-Co | 4.93 | 4.02 | 2.846 | 7.16 | 2.968 | 1.04 |
| Ce-Ni | 4.88 | 4.01 | 2.817 | 7.22 | 2.993 | 1.06 |
| Ce-Pt | 5.37 | 4.38 | 3.100 | 7.73 | 3.205 | 1.03 |
| Dy-Co | 4.93 | 3.99 | 2.846 | 7.19 | 2.981 | 1.05 |
| Dy-Ni | 4.87 | 3.97 | 2.812 | 7.16 | 2.968 | 1.06 |
| Dy-Rh | 5.14 | 4.29 | 2.968 | 7.49 | 3.105 | 1.05 |
| Er-Co | 4.89 | 4.00 | 2.820 | 7.16 | 2.966 | 1.05 |
| Er-Ni | 4.86 | 3.97 | 2.804 | 7.13 | 2.955 | 1.05 |
| Er-Rh | 4.86 | 3.97 | 2.804 | 7.44 | 3.086 | 1.10 |
| Gd-Co | 4.97 | 3.97 | 2.869 | 7.25 | 3.006 | 1.05 |
| Gd-Ni | 4.91 | 3.97 | 2.835 | 7.20 | 2.985 | 1.05 |
| Gd-Rh | 5.17 | 4.31 | 2.985 | 7.56 | 3.134 | 1.05 |
| Ho-Co | 4.88 | 4.01 | 2.818 | 7.17 | 2.974 | 1.06 |
| Ho-Ni | 4.87 | 3.97 | 2.812 | 7.14 | 2.959 | 1.05 |
| La-Ir | 5.39 | 4.20 | 3.112 | 7.69 | 3.188 | 1.02 |
| La-Ni | 5.02 | 3.99 | 2.898 | 7.39 | 3.064 | 1.06 |
| La-Pt | 5.39 | 4.38 | 3.112 | 7.78 | 3.225 | 1.04 |
| Lu-Ni | 4.83 | 3.97 | 2.791 | 7.06 | 2.929 | 1.05 |
| Nd-Co | 5.02 | 3.98 | 2.898 | 7.22 | 2.993 | 1.03 |
| Nd-Ir | 5.32 | 4.33 | 3.074 | 7.60 | 3.151 | 1.03 |
| Nd-Ni | 4.93 | 3.96 | 2.846 | 7.27 | 3.014 | 1.06 |
| Nd-Pt | 5.35 | 4.39 | 3.089 | 7.69 | 3.188 | 1.03 |
| Pr-Co | 5.01 | 3.98 | 2.893 | 7.31 | 3.031 | 1.05 |
| Pr-Ni | 4.96 | 3.98 | 2.864 | 7.28 | 3.018 | 1.05 |
| Pr-Pt | 5.35 | 4.39 | 3.089 | 7.65 | 3.172 | 1.03 |
| Pu-Ni | 4.87 | 3.97 | 2.812 | 7.14 | 2.960 | 1.05 |
| Pu-Pt | 5.26 | 4.39 | 3.037 | 7.63 | 3.163 | 1.04 |
| Sc-Ni | 4.74 | 3.76 | 2.737 | 6.93 | 2.873 | 1.05 |
| Sm-Co | 5.00 | 3.96 | 2.888 | 7.27 | 3.012 | 1.04 |
| Sm-Ni | 4.93 | 3.98 | 2.844 | 7.23 | 2.997 | 1.05 |
| Sr-Pd | 5.41 | 4.42 | 3.123 | 7.82 | 3.242 | 1.04 |
| Sr-Pt | 5.39 | 4.37 | 3.112 | 7.77 | 3.221 | 1.04 |
| Th-Co | 4.95 | 3.98 | 2.856 | 7.21 | 2.989 | 1.05 |
| Tb-Ni | 4.89 | 3.97 | 2.826 | 7.16 | 2.968 | 1.05 |
| Tb-Rh | 5.13 | 4.29 | 2.964 | 7.49 | 3.106 | 1.05 |
| Th-Ir | 5.33 | 4.27 | 3.077 | 7.66 | 3.177 | 1.03 |
| Yb-Ni | 4.84 | 3.96 | 2.796 | 7.09 | 2.941 | 1.05 |
| Y-Co | 4.93 | 3.99 | 2.846 | 7.22 | 2.993 | 1.05 |
| Y-Ni | 4.88 | 3.97 | 2.817 | 7.19 | 2.981 | 1.06 |
| Y-Rh | 5.14 | 4.29 | 2.968 | 7.50 | 3.109 | 1.05 |

The shortest bond lengths in the $MgCu_2$ type and $CaCu_5$ type compounds occurring in the same binary system are compared. The ratio of the bond lengths is ≈ 1.05 on an average.

**TABLE II. The shortest bond lengths (d in Å) in the $MgCu_2$ and CsCl type compounds**

| System | CsCl type | | $MgCu_2$ type | | RATIO |
|---|---|---|---|---|---|
| | a (Å) | $d_3$ (Å) | a (Å) | $d_4$ (Å) | $d_4/d_3$ |
| Be-Ti | 2.94 | 2.546 | 6.45 | 2.674 | 1.05 |
| Ca-Pd | 3.52 | 3.048 | 7.67 | 3.180 | 1.04 |
| Dy-Rh | 3.40 | 2.944 | 7.49 | 3.105 | 1.05 |
| Er-Rh | 3.36 | 2.910 | 7.44 | 3.086 | 1.06 |
| Gd-Rh | 3.44 | 2.979 | 7.56 | 3.134 | 1.05 |
| Hf-Co | 3.16 | 2.737 | 6.91 | 2.865 | 1.05 |
| Ho-Ir | 3.38 | 2.927 | 7.49 | 3.105 | 1.06 |
| Ho-Rh | 3.38 | 2.927 | 7.43 | 3.078 | 1.05 |
| Lu-Ir | 3.32 | 2.875 | 7.45 | 3.090 | 1.07 |
| Lu-Rh | 3.33 | 2.884 | 7.42 | 3.075 | 1.07 |
| Mg-Ce | 3.91 | 3.386 | 8.73 | 3.619 | 1.07 |
| Mg-Cd | 3.82 | 3.308 | 8.55 | 3.545 | 1.07 |
| Mg-La | 3.97 | 3.438 | 8.79 | 3.644 | 1.06 |
| Mg-Nd | 3.86 | 3.343 | 8.66 | 3.590 | 1.07 |
| Mg-Pr | 3.88 | 3.360 | 8.69 | 3.603 | 1.07 |
| Mg-Sm | 3.85 | 3.334 | 8.62 | 3.574 | 1.07 |
| Pu-Ru | 3.36 | 2.910 | 7.48 | 3.101 | 1.07 |
| Sc-Al | 3.45 | 2.988 | 7.58 | 3.143 | 1.05 |
| Sc-Co | 3.15 | 2.728 | 6.92 | 2.869 | 1.05 |
| Sc-Ir | 3.21 | 2.776 | 7.35 | 3.046 | 1.10 |
| Sc-Ni | 3.17 | 2.745 | 6.93 | 2.873 | 1.05 |
| Sm-Rh | 3.47 | 3.005 | 7.54 | 3.126 | 1.04 |
| Tb-Rh | 3.42 | 2.962 | 7.49 | 3.106 | 1.05 |
| Ti-Co | 2.99 | 2.590 | 6.69 | 2.774 | 1.07 |
| Tm-Ir | 3.35 | 2.901 | 7.48 | 3.100 | 1.07 |
| Tm-Rh | 3.36 | 2.910 | 7.42 | 3.075 | 1.06 |
| Yb-Ir | 3.35 | 2.901 | 7.48 | 3.099 | 1.07 |
| Yb-Rh | 3.35 | 2.901 | 7.43 | 3.081 | 1.06 |
| Y-Ir | 3.41 | 2.953 | 7.52 | 3.118 | 1.06 |
| Y-Rh | 3.41 | 2.953 | 7.50 | 3.109 | 1.05 |
| Zr-Co | 3.19 | 2.763 | 6.95 | 2.881 | 1.04 |
| Zr-Ir | 3.32 | 2.875 | 7.35 | 3.047 | 1.06 |
| Zr-Zn | 3.34 | 2.893 | 7.39 | 3.064 | 1.06 |

The shortest bond lengths in the $MgCu_2$ type and CsCl type compounds occurring in the same binary system are compared. The ratio of the bond lengths is $\approx 1.06$ on an average.